\documentstyle[emulateapj,psfig]{article}

\makeatletter

\newenvironment{inlinefigure}{%
\def\@captype{figure}%
\noindent\begin{minipage}{0.999\linewidth}\begin{center}}
{\end{center}\end{minipage}\smallskip}
\makeatother

\newcommand{\mum}{$\,\mu$m}
  \def\cle      {{$_ <\atop{^\sim}$}}
  \def\cge      {{$_ >\atop{^\sim}$}}

\lefthead{Chapman et al.}

\begin{document}
\title{Westphal-MMD11: An interacting, submillimeter luminous Lyman break galaxy}
\author{S.\,C.\ Chapman,$\!$\altaffilmark{1}
A.\ Shapley,$\!$\altaffilmark{2}
C.\ Steidel,$\!$\altaffilmark{2}
R.\ Windhorst,$\!$\altaffilmark{3}
}

\altaffiltext{1}{Department of Physics, California Institute of Technology, MS 320-47, Pasadena, CA, 91125}
\altaffiltext{2}{Palomar Observatory, California Institute of Technology, MS 432-10, Pasadena, CA, 91125}
\altaffiltext{3}{Arizona State University, Dept.\ of Physics and Astronomy,
Tempe, AZ, 85287-1504}
\slugcomment{To appear in the Astrophysical Journal Letters}

\begin{abstract}
We present new {\it Hubble Space Telescope}, 
high-resolution optical imaging of the 
submm luminous Lyman-break galaxy, Westphal-MMD11, an interacting
starburst at $z=2.979$. 
The new imaging data, in conjunction with re-analysis of Keck optical and
near-IR spectra, demonstrate MMD11 to be an interacting
system of at least three components: a luminous blue source, a fainter blue
source, and an extremely red object (ERO) with $R-K$\cge 6. 
The separations between components are $\sim$8~kpc ($\Lambda=0.7$, $\Omega_M=0.3$, 
h=0.65), similar to 
some of the local ultra-luminous infrared galaxies (ULIGs).
The lack of obvious AGN in MMD11, along with the fragmented, early stage merger 
morphology, suggest a young forming environment.
While we cannot unambiguously identify the location of the far-IR
emission within the system, analogy to similar ULIGs suggests the
ERO as the likely far-IR source.
The $>$10$^{12}$\,L$_\odot$ bolometric luminosity of MMD11 can be 
predicted reasonably from its
rest frame UV properties once all components are taken into account, 
however this is not typically the case for local galaxies
of similar luminosities. 
While LBGs as red in $g-R$ and $R-K$ as MMD11 are rare, they can only be
found over the restricted $2.7 < z < 3.0$ range.
Therefore a substantial number of MMD11-{\it like} galaxies 
(\cle 0.62\,arcmin$ ^{-2}$) may exist when
integrated over the likely redshift range of SCUBA sources ($z=1
-5$), suggesting that SCUBA sources should not necessarily be
seen as completely orthogonal to optically selected galaxies.
\end{abstract}


\section{Introduction}
\label{secintro}

Our understanding of the blank field submillimeter (submm) sources,
their diversity, and
their connection to optically selected galaxies has had a slow 
progression since their discovery by Smail, Ivison \& Blain (1997).  
The few sources with well studied properties come mostly from 
the lensed surveys of Smail et al.~(2002), and have been restricted
to the small fraction with relatively bright and obvious optical counterparts.
Only one of these does
not contain the signatures of an AGN (SMM\,J14011+0252 -- 
Ivison et al.~2001, 2000). 
While the identification of submm sources has thus far only been 
successful for the brightest
optical counterparts and the strongest emission line spectra,
none of the emission lines 
are all that strong compared to purely star forming objects, and it is
as yet unclear what the relative proportion of AGN is for submm sources. 
This bright counterpart 
bias also manifests itself in the redshift distribution of submm 
sources. While photometric studies place the median redshift for the
submm sample at $<z>\sim3$ (Smail et al.~2000, Archibald et al.~2001), 
no submm selected galaxy has yet been 
spectroscopically identified with redshift higher than $2.8$.

Westphal-MMD11 (hereafter MMD11) 
remains amongst the highest spectroscopically
confirmed redshift submm sources
($z=2.979$) which are thought to be exclusively starbursting galaxies.
Targeted submm observations of rare, luminous AGN at $z>3$ have
however yielded
detections since early single bolometer observations 
(Hughes et al.~1997, summarize the state of submm detections with pre-SCUBA
instruments; Archibald et al.~2001, Carilli et al.~2001, Willott et al.~2001
have demonstrated the high frequency of high$-z$ AGN detection in the submm).
Ironically, MMD11 was discovered originally, 
not through its copious submm 
emission, but as a Lyman-break galaxy (LBG -- 
Steidel, Pettini \& Hamilton 1995).
MMD11 was isolated as a candidate LBG with very high star formation
rate (SFR), its rest frame UV properties suggesting several hundred 
M$_\odot$/yr, and its $R-K=4.54$ is the reddest 
color for the observed LBG sample
(Shapley et al.~2001). The submm detection
validated the UV-based predictions, 
showing that this is an ultraluminous IR galaxy (ULIG)
with a far-IR luminosity of $L_{\rm FIR}\sim6.6\times10^{12}$\,L$_\odot$,
implying a SFR $\sim$10$^3$\,M$_\odot$/yr.
The detailed properties of this source have prompted successful submm 
detection of other LBGs (Chapman et al.~2000b).

However, LBGs have been difficult to consistently detect in the submm.
Work by Chapman et al.\ (2000a), Adelberger \& Steidel (2000),
Peacock et al.\ (2000), Eales et al.\ (2000), van der Werf et al.\ 
(2000), Sawicki (2001), Baker et al.\ (2001), Webb et al.\ (2002) has suggested 
that the direct overlap between the SCUBA and LBG populations is small and
that potential overlap may be difficult to predict from knowledge of UV
characteristics. Therefore, an 
unresolved issue is the overlap between submm selected and optically 
selected galaxy populations, with  
MMD11 remaining an enigmatic bridge.
Attempts to model the properties of 
MMD11 led to apparently contradictory pictures. 
The galaxy shows 
an unremarkable optical spectrum for a ULIG with three times the
bolometric luminosity of Arp220 (Chapman et al.~2000a). 
In the near-IR, it shows narrower nebular line widths than the median LBG
(Pettini et al.~2001), but is not unusual in the line ratio diagnostics.
MMD11 has broad-band colors that cannot be simultaneously fit in both the
UV/optical and near-IR with synthetic spectral templates; adding sufficient
dust to the template to match the $J$, $H$, $K_s$ photometry produces an
increased extinction at shorter wavelengths resulting in an
underestimate of the UV/optical bands.
The broad band photometry is therefore best fit by the superposition of
a blue, relatively unobscured galaxy coupled with a young, 
dusty galaxy (Shapley et al.~2001). 

In this paper we present new {\it Hubble Space Telescope} optical imaging
of MMD11, using existing Keck spectroscopy and near-IR imaging to 
elucidate the high resolution HST data. 
All calculations assume a $\Lambda=0.7$, $\Omega_M=0.3$ cosmology, with
h=0.65 providing a scale of 1\arcsec\ = 8.31\,kpc at $z=2.979$.

\section{Observations and Results}
\subsection{HST-Visible and Keck-NearIR observations}

HST imaging was obtained through a Cycle\,10 program with the
{\it Space Telescope Imaging Spectrograph} (STIS) to study the
morphologies of submm luminous galaxies. One orbit of integration time,
giving 1170\,sec of {\it LOW SKY} observation,
was split between two exposures, using the $50CCD$-clear filter. The
pipeline processed frames were calibrated, aligned, and cosmic ray rejected,
using standard IRAF/STSDAS routines. 
The pixel size in the STIS image is 0.0508\arcsec/pixel. The sensitivity
limit reached is $50CCD\sim27$ (5$\sigma$), 
corresponding to $R\sim28$ for a point
source with a late-type spiral galaxy SED.
The $50CCD$-clear filter is roughly a gaussian with 1840\AA\ halfwidth
and a pivot wavelength of 5733.3\AA, and we refer to the associated AB
magnitude as $R'(573)$.
Near-IR observations in 0.4\arcsec\ $K$-band seeing 
were obtained with the Near-InfraRed Camera (NIRC) on 
the Keck\,I telescope in the $J$, $H$, and $K_s$-bands, as described
originally in Shapley et al.~(2001).
The STIS image is presented in Fig.~1a with $K_s$-band contours overlaid.
The $K_s$ limiting magnitude in a 1.5\arcsec\ aperture is 22.1, 5$\sigma$.

The relative astrometry was ascertained in the images by smoothing the STIS
image to the $K_s$ resolution and matching all
sources $>5\sigma$ except MMD11. 
After, maximimizing the cross-correlation signal 
between frames, the r.m.s. of the match between the optical and infrared
sources is 0.16\arcsec.
While ground based optical imaging with $\sim1$\arcsec\ seeing identified
only an $R_s=24.05$ unresolved source (Shapley et al.~2001), our STIS imaging
uncovers three distinct components with intervening low surface brightness
emission in the MMD11 region, identified
as $B1$, $B2$, and $R3$. The component separations are: 
$B1-B2$=1.13\arcsec, $B1-R3$=0.71\arcsec, $B2-R3$=0.74\arcsec.
A 2.5\arcsec\ aperture subsuming all three components and intervening structure
measures $R'(573)$=24.03, whereas the sum of the three component measurements
in 0.4\arcsec\ apertures amounts to $R'(573)$=24.54, or 63\% of the total
flux ($R_{B1}=25.28$, $R_{B1}=26.11$, $R_{R3}=25.75$). We therefore see a
significant portion of the total flux from MMD11 is emitted in low 
surface brightness regions between the compact components.
Component $B1$ has close to half the $R'(573)$=24.54 measurement for the
triplet, with $B2$ and $R3$ having respectively 1/6 and 1/3.
The R3 component is resolved by STIS, but is quite compact
and isolated with a half light
radius of 0.2\arcsec. B1 on the other hand is morphologically complex,
with low surface brightness emission extending to relatively  
large radius.

The near-IR images of MMD11 are all nearly unresolved, and of 
high enough resolution to identify their location amongst the STIS components.
The peaks of the 
bright $J$, $H$, and $K_{\rm s}$ 
sources all align to within 0.2\arcsec\ with the
component labeled $R3$, and all 3 bands show an extension towards the 
STIS-identified blue component, $B1$.
The $K_s$-band image has the best seeing ($\sim0.45$\arcsec), and a faint
source is visible at the position of $B1$. A point source fit to the bright
R3 component is subtracted (Fig.~1b) and an
aperture measurement at $B1$ with with matched 0.5\arcsec\ diameter shows 
$K_s=21.9$,
suggesting a plausible infrared counterpart to the bright optical LBG
with $R-K=2.7$, close to the median for the LBG sample of $R-K=2.85$
(Shapley et al.~2001).	
We conclude that the 
optically bright component and the 
near-IR bright component (Shapley et al.~2001) are two distinct
parts of an interacting system. 
We note that MMD11 displays a striking difference between its rest-frame
UV and Visible emission (a {\it morphological k-correction}), 
seen only rarely in more local galaxies (Hibbard \& Vacca (1997)
Abraham et al.\ (1999), Kuchinski et al., 2001, Windhorst et al.~2002).

\subsection{Spectroscopic observations}
\label{secspec}

Keck, Low Resolution Imaging Spectrograph (LRIS) and NIRSPEC observations were obtained for MMD11.
By chance, the slit alignments on both LRIS and NIRSPEC were such that
we find evidence for redshifts of the 3 components labeled in Fig.~1.
The reductions of the spectra are described in 
Pettini et al.~(2001).
With NIRSPEC, the de-rotator was not functioning during the observation,
and the slit rotates in time, starting at zero degrees (North) and
rotating through to  $-30$ deg (West of North). While a source with a strong 
continuum is always present in the 2D spectra, a second source 
appears to rotate in and out of the 1\arcsec\ slit with time, having
a strong [O{\sc iii}]5007 emission line at the same redshift
as the bright $K$-band source, but no detected continuum (Fig.~2). 
The separation of the two components is $0.7$\arcsec, similar to the
separation of either $B1/R3$ or $B2/R3$ seen in the STIS image.
There remains some abiguity as to whether the second source lacking continuum
is $B1$ or $B2$, or both at different times.
Analysis of the two components in the NIRSPEC spectrum
indicates that they have roughly equal [OIII]$_{5007}$ flux.
While H$\beta$ and [OIII]$_{4959}$ are also visible in bright $K$-band source,
there are no other lines visible for the off-continuum component.
[OIII]$_{4959}$ in particular lies in a much noiser region of the $H$-band
window than [OIII]$_{5007}$, and is only just detected in the brighter 
$R3$ component.

The LRIS slit was aligned with the brightest $R$-band peak, which
we now identify as STIS-$B1$.
Close inspection of the LRIS image suggests that a spatially extended or double 
Ly$\alpha$ line lies on the slit, with an apparent extention of $\sim1$\arcsec.
The slit is aligned North/South, and the extension of the line
has no continuum associated. We conclude that the red source, STIS-$R3$, 
was also present on the slit with detected Ly$\alpha$.

\subsection{Submillimeter and Radio Observations}
\label{secsmm}

SCUBA observations at 850~$\mu$m and 450~$\mu$m
were taken during an observing runs in 1998. Reductions
and measurements from the data were presented in Chapman et al.~(2000a).
The photometry detection at 850~$\mu$m measured $5.5\pm1.4$\,mJy.
MMD11 has also been detected at 1200\mum\ at IRAM 
(A. Baker, private communication). 
VLA radio observations were obtained in A-configuration at 1.4\,GHz.
These observations were made available to us to
search for a radio counterpart to the submm source in order to 
pinpoint the location of the far-IR emission.
The reductions and details are described elsewhere
(M.\ Yun et al.~in preparation). 
No significant emission is detected at the optical source position, or within 
the SCUBA beam, and we place a 3$\sigma$=\,75$\mu$Jy limit on the 1.4\,GHz flux.

\section{Analysis and Discussion}

While early modeling efforts for MMD11 did not result in any conclusive
picture, our deep STIS imaging in conjunction with the Keck spectra
and near-IR imaging have revealed the source to be
composed of at least three distinct components (Figs.~1,2) lying 
at the same redshift.
The $R3$ component represents an ERO with $R'(573) - K_{\rm s} = 6.15$ 
generating the
bulk of the near-IR emission, and can naturally be fit with a dusty
template SED.
By subtracting a point source fit to the $K_s$-band R3, we have also measured 
the $K$ flux from B1, resulting in $R'(573) - K_{\rm s} = 2.7$, a value
fairly typical of LBGs.
However, a template SED fit to an ERO at $z\sim3$ 
requires a large amount of extinction using the Calzetti dust law
(Calzetti~1997),
as found even for the average properties of LBGs like MMD11 in
Shapley et al.~(2001). For MMD11 in particular, this would imply an 
unphysically large star formation rate, dwarfing the submm implied
600\,M$_\odot$\,yr$^{-1}$ (Chapman et al.~2000a), 
as even the ERO component $R3$ 
has 1/5 the total system flux at $R$-band (translating to an
uncorrected SFR of $\sim4$M$_\odot$\,yr$^{-1}$).
In addition, the large bandwidth of the STIS $50CCD$ filter precludes 
accurate constraints on the optical SED for the various components.
We 
simply conclude that the majority of the near-IR emission
emminates from the $R3$ component.

With such disparate broadband and morphologocial 
properties between components, it is 
surprising that the $[\rm{OIII}]_{\lambda5007}$
from different components are roughly equal in intensity
(EW$_\lambda$ is of course very different).
We assume that the similarities occur only by chance. 

While we have elucidated the optical and near-IR structure of MMD11
significantly with our high resolution imaging and spectroscopy,
we would like to identify precisely the location and nature of the intense
far-IR emission within the system.
Our radio data is not deep enough to pinpoint the far-IR emission
within MMD11, although it should have detected a buried 
AGN component if present. The upper limit (S$_{1.4 \rm{GHz}}<75\mu$Jy),
is still within the error bounds of the median expected flux relation
for the submm measurement (30\,$\mu$Jy -- using Carilli \& Yun 2000). 
The luminosity limit (2.00$\times$10$^{22}$\,W/Hz, assuming a radio
spectral index of $\alpha=-0.75$) is a factor 10 to 100 less than $z\sim0.1$
radio quiet quasars studied by Kukula et al.\ (1998). 
The LRIS spectrum of MMD11 shows no evidence for narrow (Type II AGN) or 
broad (Type I AGN) emission lines. The Ly$\alpha$ profile, 
known for both B1 and
R3, has a total equivalent width of zero considering
both the emission and absorption. The Ly$\alpha$ emission line EW$\lambda$ is
only 6\AA\ in the rest frame, which is weak compared with the 
Ly$\alpha$ emission EW$\lambda$ distribution of 
LBGs (Steidel et al.\ 1999).
For component R3, the H$\beta$ line is narrow, and the 
[OIII]/H$\beta$ ratio is consistent with a starburst. Therefore
the spectroscopic data can be used to place strong limits on 
the AGN contribution to MMD11.

While we have no way to directly assess the location or extent 
of the far-IR emission,
other ULIG systems often show
submm emission arising in the ERO galaxy component, 
while the bulk of the optical
emission remains relatively unobscured in bluer companions. 
The only submm-selected source identified with a 
pure starburst nature is SMM14011-0252. This galaxy
was recently characterized with a similar multi-component
structure (Ivison et al.~2001), encompassing an ERO separated by 
$\sim$1.5\arcsec\ from 
each of two luminous bluer sources, the far-IR emission apparently 
localized near the ERO from the radio and CO gas (Frayer et al.~1999,
Ivison et al.~2000). 
MMD11 has a similar far-IR luminosity to SMM14011-0252, which has
a cluster lensing corrected SCUBA flux of 5.0\,mJy and a redshift of $z=2.56$,
although the CO(3-2) measurement would not have been possible without the
2.5$\times$ lensing factor.
A similar multi-component configuration with the radio/far-IR identified
to an ERO is also seen in two $z$\cle 1 SCUBA-detected 
ULIGs from the ISO/FIRBACK survey (Chapman et al.~2002a).

Both MMD11 and SMM14011-0252 appear morphologically to be examples of the
{\it early stage merger} type of ULIG 
(Goldader et al.~2001), prevalent in $\sim25$\% of the local ULIG population
whereby luminous components are separated by $>10$\,kpc.
These sources are in some sense equally well
identified from their submm/far-IR or UV properties.
While this class of system appears relatively common in SCUBA
identifications, 
it is as yet unclear whether this is because these sources were the
easiest to identify of the submm population,
or because submm galaxies are typically made in different ways than
local ULIGs, such early stage mergers being much more common
at high-$z$. Recent work with HST has suggested the merger rate in the past
being higher by a factor (1+$z$)$^{2.5}$ (Le Fevre et al.~2000).

While MMD11 is clearly similar in many respects to ULIGs at
more recent cosmic epochs, it remains difficult to relate MMD11 
to the rest of the LBG population.
Most of the LBGs can be consistently fit from $U$- through $K$-band with 
a single SED model (Shapley et al.~2001).
However, the reddest $R-K$ LBGs have significant far-IR/submm emission as
a group, and another LBG with $R-K>4$ has been detected with SCUBA
(Chapman et al.~2000b). It is therefore reasonable to 
hypothesize that the most bolometrically luminous tail of the LBG 
population extends into the $z\sim3$ ULIG/SCUBA population.

When considered as a complete system (all components included), 
both MMD11 and SMMJ14011 appear to lie roughly on the local far-IR/UV versus
UV continuum slope relation for local IUE starburst galaxies
(Meurer et al.~1999 -- FIX/$\beta$ relation).
However, we have seen that the regions generating the UV and far-IR are 
likely to be physically separated by $\sim8$\,kpc, and it might be expected 
that the predictive qualities of FIX/$\beta$ are
likely to be haphazard and orientation-dependent. 
We must thus consider why 
these two $z\sim3$ galaxies balance in energy output between the 
UV and far-IR well enough for the prediction to work.
MMD11 is quite luminous, even without any extinction
correction; the SFR calculated from L$_{\rm UV}$ is 20\,M$_\odot$/yr.
In addition, the total MMD11 system is, by UV color,
very red, which explains why it was flagged as a submm-bright candidate
with a predicted S$_{\rm 850 \mu m}\sim2$\,mJy,
in the absense of knowledge about the $R-K$ color (MMD11 would not have
been selected for submm followup had it been truly unobscured).
Systems like MMD11 and SMMJ14011 may therefore suggest that an LBG with 
a luminous, relatively red, UV component could imply
an associated bolometrically luminous system.
In other words, the existence of a bright, reddened
UV structure makes it causally more likely that there is a 
bolometrically bright and heavily obscured component lurking nearby.
It is not clear that either SMMJ14011 or MMD11 are necessarily {\it accidents}
just because they have multiple components with different amounts of UV
attenuation.
In this context, the FIX/$\beta$ relation can be interpreted as mapping
bolometrically luminous galaxies to dustier configurations,
and even the parts that are able to leak out in the UV will tend
to be reddened more so than if the object were dust-free.

The FIX/$\beta$ relation holds for the local LIG/starburst
population (L$_{\rm FIR} \leq 10^{11}$\,L$_\odot$),
despite the UV and far-IR bright components often being spatially distinct.
By contrast, this is
not a typical property of ULIGs ($>10\times$ more far-IR luminous 
than the LIGs and IUE starbursts), or even the
early stage merger subset of the ULIGs, which are generally many times
over-luminous in the far-IR compared to the predicted UV energy absorbed. 
From the ULIG sample
of Goldader et al.~(2002), only one other galaxy, IRAS\,22491,
appears to obey the FIX/$\beta$ relation. 
However, the $z\sim3$ LBGs have greater characteristic 
luminosities for a given farIR/UV ratio than local galaxies,  
suggesting that LBGs are less obscured per unit luminosity than lower redshift
starforming galaxies (Adelberger \& Steidel 2000, Chapman et al.~2000b).
This suggests that LBGs are typically more similar to the less luminous 
LIG/starburst population locally than the ULIGs.

LBGs as red as MMD11, in both $g-R$ and $R-K$, are rare
($<$5\%, or $<$0.05\,arcmin$^{-2}$).
However, the statistics for such sources
are poor since the LBG survey could only have found objects like
MMD11 (i.e., instrinsically red in the UV) over a very restricted redshift
range, $\sim 2.7 < z < 3.0$
(at higher redshifts MMD11 would have been too red in $g-R$
to have remained in the sample).
Therefore, when integrated over the likely redshift range of SCUBA sources ($z=1
-5$),
there may be a substantial number of MMD11-{\it like} galaxies ($<$0.62\,arcmin$
^{-2}$
for our adopted $\Lambda$-cosmology),
suggesting that SCUBA sources should not necessarily be
seen as completely orthogonal to optically selected galaxies.
While the conclusions extrapolated from MMD11 may be a misleading effect of
small number statistics, the finding is not necessarily 
inconsistent with the
current state of followup to blank field and radio-identified submm sources
(e.g., Smail et al.~2002, Chapman et al.~2002b, Ivison et al.~2002).
These studies 
reveal a broad range in optical properties for the SCUBA population.
At least in part this may be explained because
some of the submm galaxies
comprise both obscured and unobscured interacting components like
MMD11 and SMM14011-0252.

\acknowledgements
We thank M.Yun for access to the VLA radio data.
An anonymous referee has helped to improve the text.
We gratefully acknowledge support from NASA through
HST grant 9174 (SCC, RW), 
awarded by the Space Telescope Science Institute. 
CCS and AES have been supported by grants
AST95-96229 and AST-0070773 from the U.S. 
National Science Foundation and by the
David and Lucile Packard Foundation.

\newpage
%
% Figure 1
%
%\begin{inlinefigure}
\begin{figure*}[htb]
\centerline{
\psfig{figure=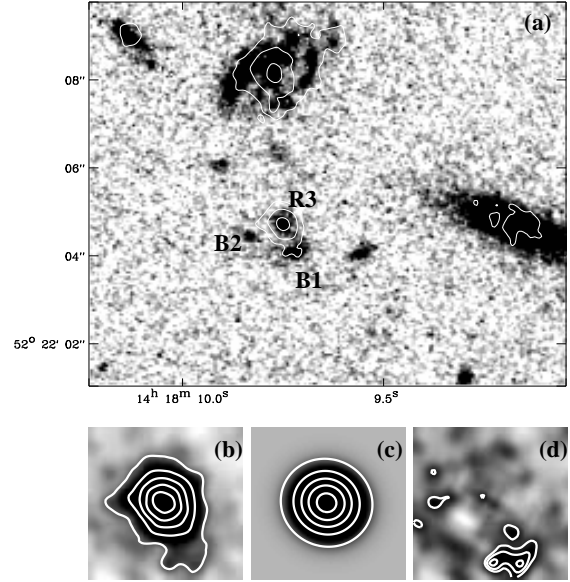,angle=-90,width=6.8in}
}
\vspace{6pt}
\figurenum{1}
\caption{
MMD11 observed with STIS $50CCD$ (greyscale) and 
$K_s$-band (Keck/NIRC) in contours. North is up, East is left, with a 9arcsec\
by 11\arcsec\ frame centered near the red R3 component. Note the faint
$K_s$ extension from R3 to B1. In the lower three panels we present a close
up view of the red R3 component (a), a gaussian fit to R3 (b),
and the residuals after subtraction where contours are 2,3,4$\sigma$ (c). 
A faint $K$ source is
present at the position of the blue source, B1.
}
\label{fig1}
\addtolength{\baselineskip}{10pt}
%\end{inlinefigure}
\end{figure*}

%
% Figure 2
%
\begin{inlinefigure}
\vspace{6pt}
\centerline{
\psfig{figure=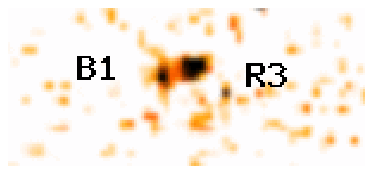,angle=0,width=3.1in}
}
\vspace{5pt}
\figurenum{2}
\caption{
Spatially extended spectra of the [O{\sc iii}]5007 line
from NIRSPEC, identifying
redshifts for 2 of the 3 components labeled in Fig.~1, STIS-$B1$,$R3$
(separation $\sim0.7$\arcsec). The image has been smoothed with a
0.4\arcsec\ gaussian.
The dispersion direction is vertical.
}
\label{fig2}
\addtolength{\baselineskip}{10pt}
\end{inlinefigure}

\end{document}